\documentclass[12pt,draftclsnofoot,onecolumn]{IEEEtran}
\usepackage[USenglish]{babel}
\usepackage{units}
\ifx\pdfoutput\undefined
  \usepackage[dvips]{graphicx}
\else
  \usepackage[pdftex]{graphicx}
\fi

\begin{document}

\title{Cooper Pair Transport in a Resistor-Biased Josephson Junction Array}
\author{Sergey~V.~Lotkhov,
        Vladimir~A.~Krupenin,
        and Alexander~B.~Zorin}
\thanks{Manuscript received July 10, 2006.
        This work was partially supported by the European Commission within the project EuroSQIP.}% <-this % stops a space
\thanks{S.~V.~Lotkhov and A.~B.~Zorin are with Physikalisch-Technische Bundesanstalt, D-38116 Braunschweig, Germany.}
\thanks{V.~A.~Krupenin is with Laboratory of Cryoelectronics, Moscow State University, 119899 Moscow, Russian Federation.}

\maketitle

\begin{abstract} The dc transport properties of long arrays of small Al Josephson junctions, biased
through on-chip Cr resistors, are studied. The $IV$-characteristics
show a large Coulomb threshold for current as well as negative-slope
regions indicating the regime of autonomous Bloch oscillations up to
rather high frequencies of $f = I/2e \sim \unit[1]{GHz}$, comparable
to those reported by other groups for single junctions. On the other
hand, a small depth of the back-bending implies a low duty cycle and
a broad spectrum of the oscillations, which we attribute to the
insufficiently high impedance of the bias resistors. A
self-sustained switching process at a small bias current is used to
study the statistics of the switching voltages and to determine the
effective Bloch capacitance which was found to considerably exceed
the geometric junction capacitance.
\end{abstract}


\IEEEpeerreviewmaketitle

\section{Introduction}

The application of single-charge tunneling effects in electrical
metrology has been considered in a view of capacitance and current
standards based on charge quantization phenomena (see, e.g., review
\cite{RevKell}). The proper charge quantization in circuits with
small tunnel junctions has been achieved for low junction
transparencies: $R_{\rm T} \gg R_{\rm Q} \equiv h/e^2 \approx
\unit[25.8]{k\Omega}$ for the single-electron tunnelling resistance,
and $E_{\rm J} \ll E_{\rm C}$ for Cooper pair devices, with $E_{\rm
J} \propto R_{\rm T}^{-1}$ and $E_{\rm C}$ being the Josephson
coupling and charging energies, respectively. For example,
remarkable advances have been made in development towards the
single-electron standard of capacitance \cite{SciKell}. The accurate
operation of these circuits, based on pumping of single electrons,
has been, however, limited to small currents, $I \sim 1$ to
$\unit[10]{pA}$, because of the stochastic nature of single electron
tunneling with a relatively long time constant.

Here we address the Cooper pair transport in the case of substantial
Josephson coupling $E_{\rm J} \ge E_{\rm C}$, i.e., when both the
Josephson phase and the charge exhibit significant quantum
uncertainties. In such rather transparent, and therefore
large-current-capable junctions, high accuracy could still be
achieved due to the exact $2e$-periodicity of their energy bands as
a function of the quasicharge \cite{LiZo}.

\section{Basic idea and sample}

The related transport phenomenon is known as Bloch oscillations in a
current-biased junction with a fundamental current-frequency
relation $I = 2ef$ \cite{LiZo}. Successful attempts to phase-lock
the Bloch oscillation by an external signal were made in the early
1990s (see, e.g., \cite{Kuzm1,Havi,Kuzm2}) in the single junctions
biased through high-ohmic on-chip microresistors for frequencies up
to $f \sim \unit[10]{GHz}$. Unfortunately, the linewidth of
oscillations was found to be rather large, $\delta f \sim
\unit[1]{GHz}$, which was supposedly due to significant thermal
fluctuations which smeared the current plateaus. Moreover, a
possible size of these plateaus was expected to be rather small,
scaling as a small threshold voltage in a single junction.

Our approach is based on 1D arrays of small Josephson junctions. As
was shown theoretically for current-biased arrays \cite{AL} and
proved experimentally for very long voltage-biased Al arrays ($N
\sim 200$) \cite{AAH}, these arrays are analogous to long Josephson
junctions: Similar to the fluxons formed in the long junctions, the
charge profile in the array is governed by the sine-Gordon type of
equation, describing the Cooper pair solitons of size (in number of
junctions) $M \sim \sqrt {\tilde C/C_{\rm 0}} < N$ \cite{AAH}. Here
we denote as $\tilde C$ the effective junction capacitance, related
to the curvature of the ground state energy $E(q)$ and, hence, the
shape of this 2$e$-soliton; $C_0 \ll \tilde C$ is the
self-capacitance of the intermediate islands. The coherent motion of
a train of 2$e$-solitons along an array biased via high-ohmic
resistors can be described by a set of non-stationary equations:

\begin{equation}
\label{SineGordon} L_B (q_i)\frac{\partial ^2 q_i }{\partial t^2} +
\frac{\partial E(q_i )}{\partial q_i } = \frac{1}{C_0}(q_{i - 1} -
2q_i  + q_{i + 1} ), \quad i = 1,2,...
\end{equation}
which includes the quasicharge-dependent Bloch inductances of the
junctions $L_{\rm B}(q_{\rm i})$ which were recently introduced in
\cite{Z}, with corresponding initial and boundary conditions. Here
we denote the set of junction quasicharges as $\{q_{i}(t)\}$. In the
limit of small Josephson coupling, $\lambda \equiv E_{\rm J}/E_{\rm
C} \ll 1$, the periodic term $\partial E(q_i )/\partial q_i $ takes
the form of the sawtooth function with the amplitude $e/C$, where
$C$ is the capacitance of the individual junctions in the array. In
the opposite case, $\lambda \gg 1$, there is an analytic expression
\cite{LiZo}:

\begin{equation}
\label{LargeLambda} \frac{\partial E(q_i)}{\partial q_i }
=a(\lambda)\frac{e}{C}{\rm sin}(\pi q_i/e),
\end{equation}
where $a(\lambda) = 2^{11/4} \pi ^{1/2} \lambda ^{3/4} \exp \left[ {
- \left( {8\lambda } \right)^{1/2} } \right]$ is a numerical factor
describing exponential suppression of the Bloch band width for
$\lambda \to \infty$.

\begin{figure}[t]
\centering%
\includegraphics[width=\columnwidth]{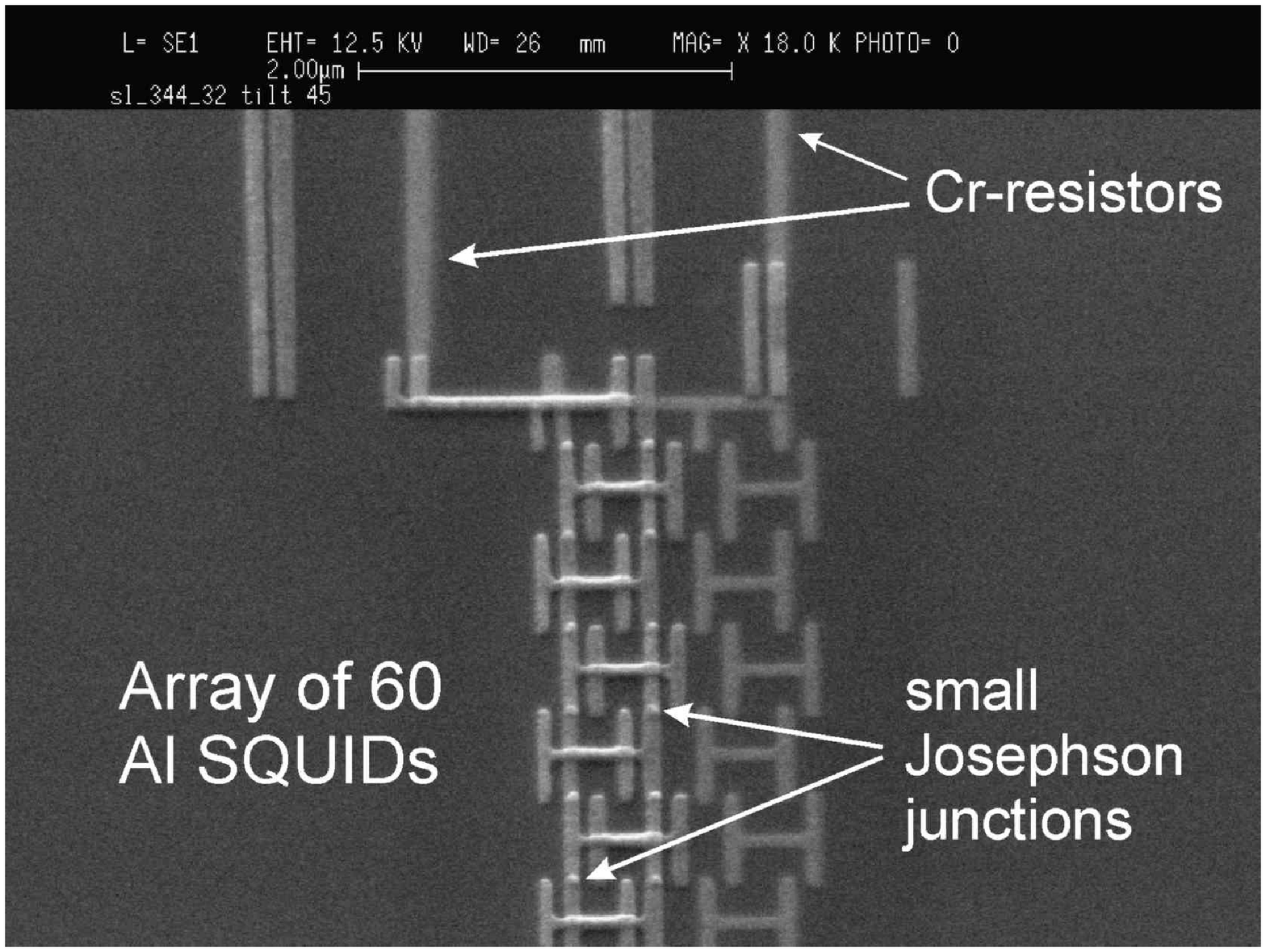}
\caption{} \label{SEM}
\end{figure}

In this paper, we report the dc properties of autonomous arrays of
$N =60$ SQUIDs (see the SEM photo in Fig.~\ref{SEM}), fabricated
using the Al/oxide/Al-Cr shadow evaporation process (see, e.g.,
\cite{Napoli}) in a 4-point layout with Cr microstrips of
$\unit[50]{\mu m}$-length with $R_{\rm Cr} \approx
\unit[550]{k\Omega}$ each. The tunnel resistance was $R_{\rm T}
\approx \unit[8]{k\Omega}$ and an estimated capacitance was $C_{\rm
T} \sim \unit[0.36]{fF}$ per junction, which corresponds to the
maximum Josephson coupling energy of each SQUID, $E_{\rm J} \approx
\unit[160]{\mu eV}$ and the charging energy $E_{\rm C} \equiv
e^2/4C_{\rm T} \approx \unit[110]{\mu eV}$ per link, yielding their
ratio $\lambda \approx 1.5$. We roughly estimate $C_{\rm 0} \sim
\unit[50]{aF}$.

The samples were measured in the current-bias mode with an external
high-ohmic bias resistor, $R_{\rm B} = \unit[100]{M\Omega}$. The
capacitance of each connection line to ground was $C_{\rm L} \sim
\unit[1]{nF}$, resulting in an $RC$-constant of the bias circuitry
of $\tau_{\rm D} \sim R_{\rm B}C_{\rm L} = \unit[0.1]{s}$. Our
microwave-tight holder was supplied with
Thermocoax$^{\textregistered}$ filters (jacket $\O~\unit[0.35]{mm}$,
110~cm long), anchored to the mixing chamber.

\section{Results}

\begin{figure}[t]
\centering%
\includegraphics[width=\columnwidth]{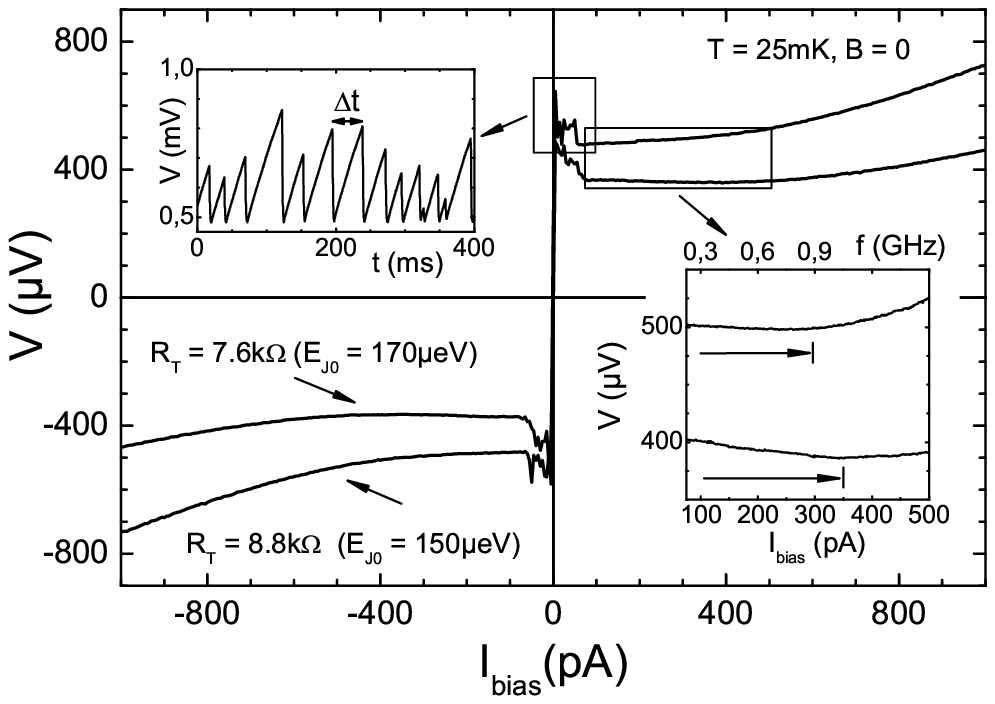}
\caption{ } \label{IVC}
\end{figure}

Several important peculiarities can be observed in the
$IV$-characteristics of the arrays shown in Fig.~\ref{IVC}. In our
opinion, the most remarkable transport property of the system
relates to the bias current range $I_{\rm bias} \le 300$ to
$\unit[350]{pA}$, with zero to negative slope of the $IV$-curve (see
the blow-up of the corresponding region in the right inset in
Fig.~\ref{IVC}), which we assume to be an indication of the Bloch
oscillation regime \cite{LiZo}. Due to the moderate current values,
the ramping of quasicharge in this transport regime occurs
sufficiently slowly that the system always stays in the zero Bloch
band and experiences 2$e$-periodic oscillations of the voltage with
the frequencies up to $f = I_{\rm bias}/2e \sim \unit[1]{GHz}$. At
larger currents, the average voltage across the array gradually
increases, presumably due to excitation of the upper energy bands
\cite{LiZo}, thus setting a high-frequency limit for possible
 application of these particular Josephson arrays as sources of a
quantized current. Note that, despite a large number of junctions in
our arrays, this limitation, being at the level of GHz, is virtually
of the same scale with the frequencies of phase-locking experiments
on single junctions \cite{Kuzm1}, whereas the typical voltage scale
is much larger, which is promising for a better observability of the
Bloch oscillations.

On the other hand, small depth of the back-bending, typically, few
percent, points out the low duty cycle of Bloch oscillations with
the short pulse duration of $\tau_{\rm arr} \le R_{\rm Cr}C_{\rm
arr} \sim (\unit[5]{GHz})^{-1} \ll f^{-1}$, where we use a rough
estimate for the "input" capacitance of the array, $C_{\rm arr}
\approx (2\tilde C C_{\rm 0})^{1/2} \approx \unit[0.4]{fF}$ (see
below for our estimation of the effective value $\tilde C$). We
attribute the present form of oscillations to insufficiently high
impedance of the biasing resistors. Simple estimations show that the
observed upper frequency limit for a single-band behaviour of about
$\unit[1]{GHz}$ is achieved by the system biased only slightly above
its Coulomb blockade threshold, therefore resulting in a typical
decay time profile of oscillations. A broad spectrum of such
oscillations, with considerable contribution of higher harmonics,
makes it difficult to ensure their effective phase-locking to an
external high-frequency signal. In particular, the preliminary
experiments on microwave irradiation of our resistor-biased arrays
up to frequencies of about $\unit[1]{GHz}$ did not result in
observable features in the $IV$-curves. On the other hand, it was
shown in \cite{Sawdrive} for the Josephson oscillations that even
such oscillations may be effectively phase-locked using a driving
signal of an appropriate waveform.

\begin{figure}[t]
\centering%
\includegraphics[width=\columnwidth]{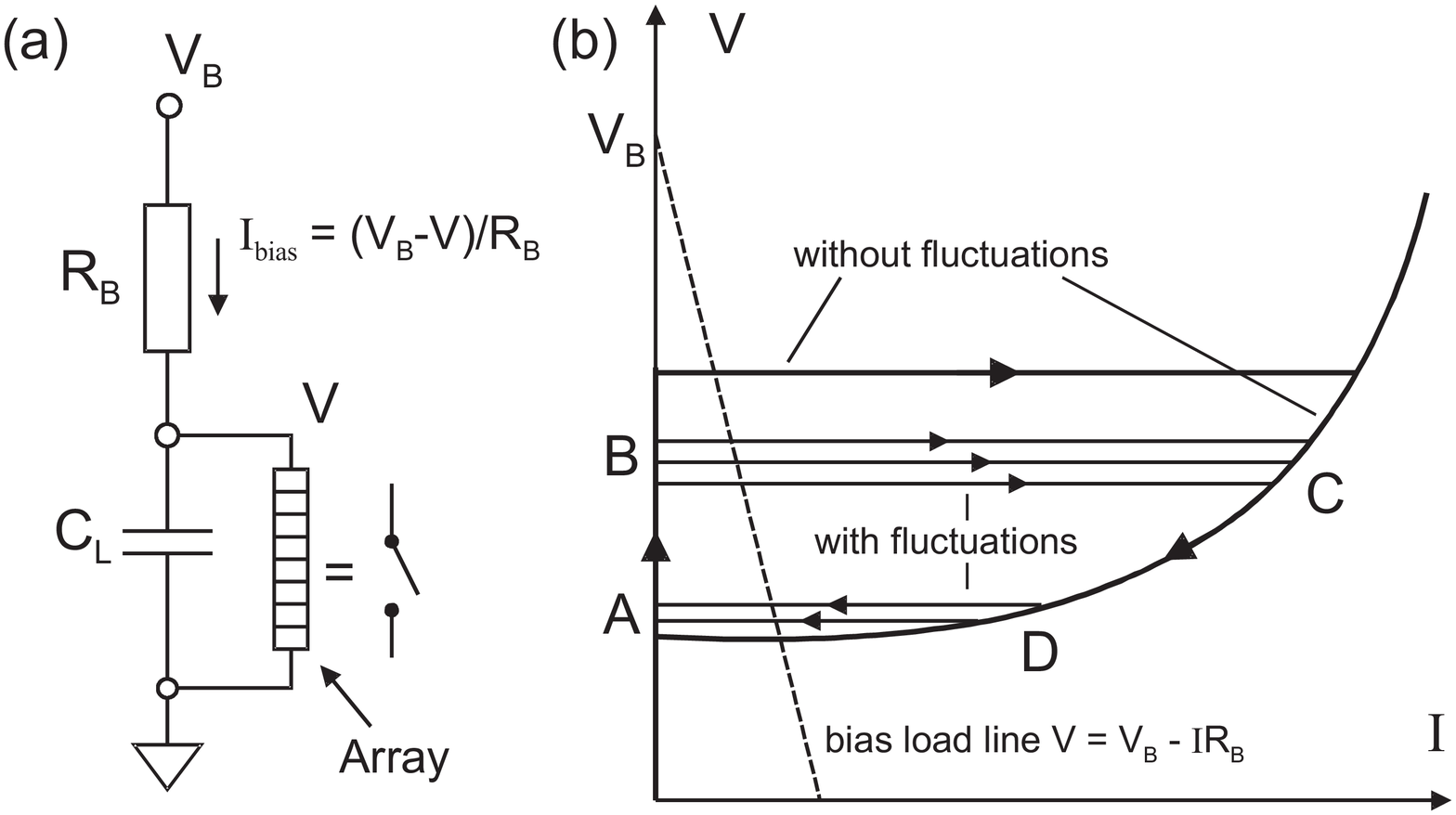}
\caption{ } \label{hystere}
\end{figure}

As an indication of a hysteretical behaviour due to the inductance
term in the equation of motion (\ref{SineGordon}) \cite{AAH,Z}, the
biasing point at small currents $I_{\rm bias} \sim \unit[10]{pA}$
was found to be unstable, exhibiting relatively slow irregular
voltage oscillations, see left inset in Fig.~\ref{IVC}. As shown
schematically in Fig.~\ref{hystere}, random switching of the array
from the blockade to the finite-current state was followed by a
relatively slow recharging of the line capacitance $C_{\rm L}$,
which made possible the real-time observation of the switching
cycles. In the blockade state, A$\to$B, with a life-time $\Delta t
\sim \tau_{\rm D}$, the current source charged the input terminal of
the array until its switching, B$\to$C, to the finite-current state,
$I_{\rm bias} \sim \unit[1]{nA}$, followed by a rapid discharging,
C$\to$D, down to voltages $V_{\rm r} \approx \unit[500]{\mu V}$ and
retrapping, D$\to$A, facilitated by the presence of fluctuations
(cf. the thermally enhanced retrapping process in a shunted
Josephson junction \cite{Ben-Jacob}).

\begin{figure}[t]
\centering%
\includegraphics[width=\columnwidth]{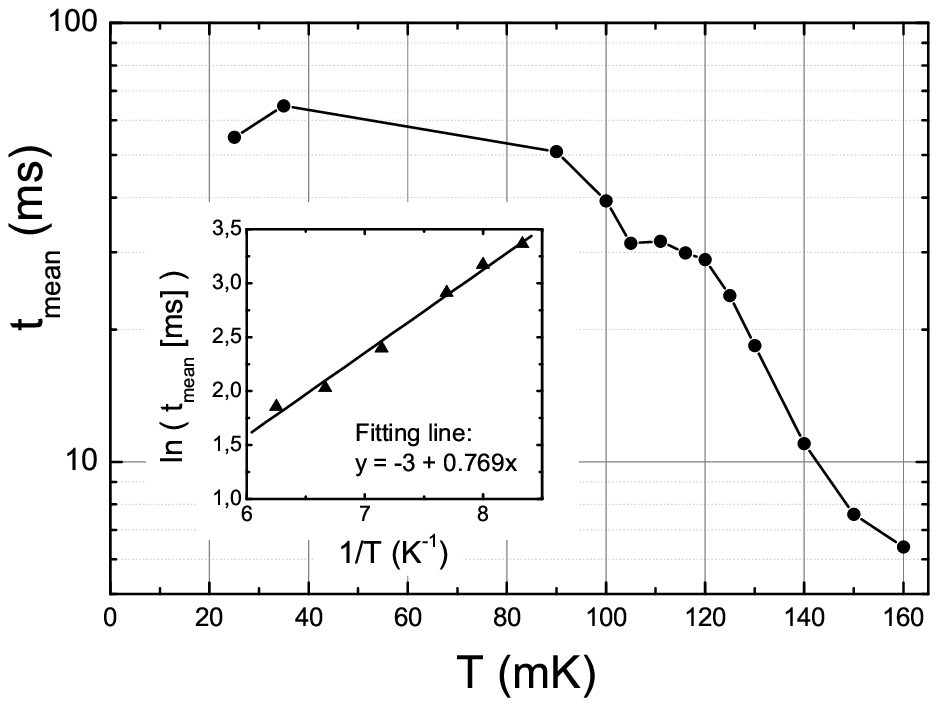}
\caption{} \label{Times}
\end{figure}

Whilst at $T < \unit[100]{mK}$ the average escape time, $t_{\rm
mean} \equiv \left\langle {\Delta t} \right\rangle$, was nearly
temperature-independent (cf. \cite{Escape}), as shown in
Fig.~\ref{Times}, the switchings at $T > \unit[100]{mK}$ were due to
thermal escape over the barrier whose height we have evaluated to be
$\Delta U \approx \unit[70]{\mu eV}$, as estimated from the slope of
the Ahrrenius plot, see inset to Fig.~\ref{Times}.

Using the approach \cite{Kazacha} which is valid, strictly speaking,
only in the weak-coupling case $\lambda \ll 1$, we obtained a rough
estimate of the effective junction capacitance and the length of a
soliton in the array. It is possible to express the barrier height
$\Delta U(V)$ through the threshold voltage $V_{\rm T}$ of soliton
motion:

\begin{equation}
\label{DeltaU} \Delta U(V) \approx \frac{eV_{\rm T}}{2}\left( {1 -
\frac{V}{V_{\rm T}}} \right)^2.
\end{equation}

\begin{figure}[t]
\centering%
\includegraphics[width=\columnwidth]{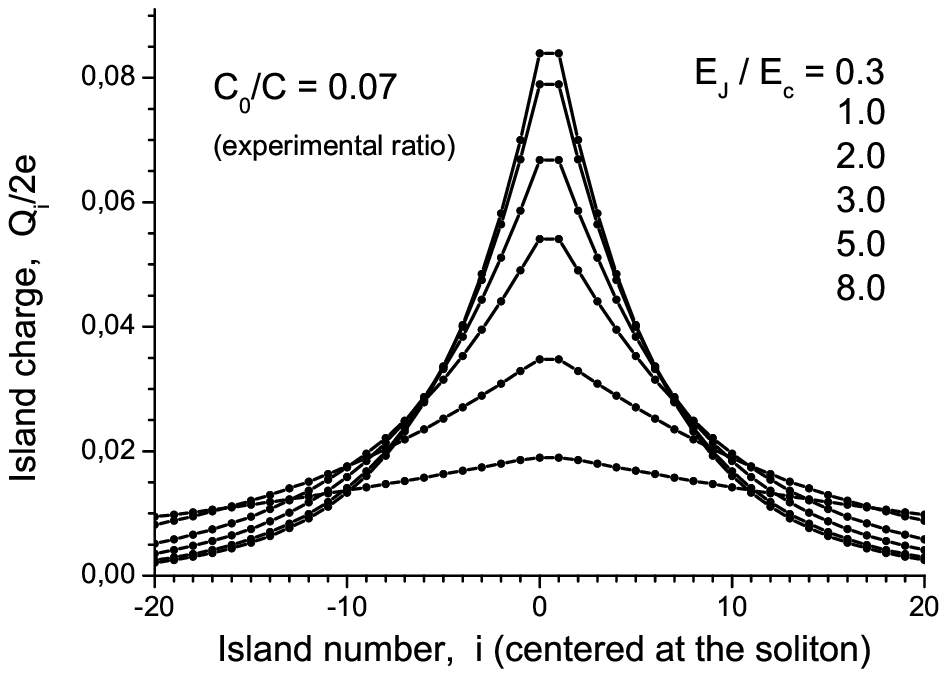}
\caption{} \label{Soliton}
\end{figure}

Taking advantage of the narrow range of switching voltages at $T >
\unit[100]{mK}$, $V \approx V_{\rm r}$, we estimated $V_{\rm T}
\approx \unit[830]{\mu V}$, a soliton energy $E \equiv \Delta U(0) =
eV_{\rm T}/2 = e^2/(2\tilde C C_{\rm 0})^{1/2} \approx
\unit[420]{\mu eV}
> E_{\rm C}$, the effective Bloch capacitance $\tilde C \sim
\unit[1.6]{fF} \gg C_{\rm T}$, and the size of a soliton $M \sim 8$.
One can see that the effective Bloch capacitance greatly exceeds the
geometric junction capacitance which is obviously due to the strong
suppression of the Bloch band width, $\Delta E = E^{\rm max}(q) -
E^{\rm min}(q)$, for the values of $\lambda >1$ [see, e.g., the
limiting case, $\lambda \gg 1$, described by equation
(\ref{LargeLambda})]. The shape of a static 2$e$-soliton in the
infinitely long array is shown in Fig.~\ref{Soliton} as calculated
by numerically solving the equation (\ref{SineGordon}) in the
stationary case $\partial q_i /\partial t \equiv 0$, for several
representative values of the ratio $E_{\rm J}/E_{\rm C}$.

\section{Conclusions and outlook}

Our dc measurements pointed out the presence of autonomous Bloch
oscillations in a resistively biased array of small Josephson
junctions. Compared to the experiments with single Bloch junctions
reported by other groups, similar Bloch frequencies are expected,
whereas the Coulomb voltage threshold and, as a consequence, the
amplitudes of oscillations are considerably larger, making a
potential advantage of using the arrays for phase-locking
experiments. However, to enable an external synchronization of these
oscillations one should realize at least several times higher bias
impedance, possibly in form of very resistive microstrips. A certain
increase in the Bloch frequency could be expected for the junctions
with a larger Josephson energy. But since the voltage scale is also
affected by the values of $E_{\rm J}$ and $E_{\rm C}$, a detailed
optimization is still important. Using the statistics of the
spontaneous state switchings, we evaluated the effective junction
capacitance and the length of a Cooper pair soliton. These data can
be helpful for future device development. For practical
applications, the influence of the background charges in the islands
of the array on the soliton dynamics should also be investigated.

\section{Acknowledgements}
The authors would like to thank D.~V.~Averin and A.~V.~Ustinov for
helpful discussions. Technical assistance from T. Weimann (e-beam
lithography), S.~A.~Bogoslovsky (measurement setup), F.-J.~Ahlers
and V.~A.~Rogalya (software) is gratefully acknowledged. The work
was partially supported by the EU through the projects RSFQubit and
EuroSQIP.

\newpage

\section*{FIGURES}

Fig.1. SEM picture of a fragment of the resistor-biased array of
two-junction SQUIDs fabricated with the three-angle evaporation
technique.

Fig.2. The $IV$-curves of two slightly different samples. Left
inset: a typical time trace of the self-sustaining switching cycles
observed at low bias currents. Right inset: a close-up look of the
negative-slope segments in the $IV$-curves (shown by horizontal
arrows) corresponding to single-band Bloch oscillations.

Fig.3. An equivalent electrical circuit (a) and a cycle diagram in
the $IV$-plane (b), both explaining the switching/retrapping
process. For the sake of simplicity, the on-chip Cr resistors and
the four-point connection layout of the experiment are not shown.
The Josephson array is schematically represented by the hatched box.
In the presence of fluctuations (external noise, thermal
fluctuations, etc.) there is no stable bias point along the load
line, which results in the self-sustaining voltage relaxation
oscillations observed in experiment.

Fig.4. Mean lifetimes of the blockade state. Inset: Ahrrenius plot
at high temperatures.

Fig.5. Calculated shape of a 2e-soliton presented in a form of the
island charges $Q_i = q_{i+1}-q_i = C_{\rm 0}V_i$, where $q_i$ is a
quasicharge of an $i$-th junction and $V_i$ is a voltage on the
$i$-th island.

\newpage

\begin{center}
S.~V.~Lotkhov $et~al.$ "Cooper pair transport in a resistor-biased
Josephson junction array"
\end{center}
\newpage

\begin{center}
S.~V.~Lotkhov $et~al.$ "Cooper pair transport in a resistor-biased
Josephson junction array"
\end{center}
\newpage

\begin{center}
S.~V.~Lotkhov $et~al.$ "Cooper pair transport in a resistor-biased
Josephson junction array"
\end{center}
\newpage

\begin{center}
S.~V.~Lotkhov $et~al.$ "Cooper pair transport in a resistor-biased
Josephson junction array"
\end{center}
\newpage

\begin{center}
S.~V.~Lotkhov $et~al.$ "Cooper pair transport in a resistor-biased
Josephson junction array"
\end{center}
\newpage

\end{document}